\documentclass[iop]{emulateapj}		
\usepackage{epsfig}			
\usepackage{graphicx,color}		
\usepackage{amssymb}			
\usepackage{color}			
\usepackage{url}			
\usepackage{amsmath}			
\usepackage{rotating}			
\usepackage{float}			
\usepackage{textcomp}			
\usepackage{psfig}
\usepackage{epstopdf}
\usepackage{dcolumn}
\usepackage{times}
\usepackage{tabularx}
\usepackage[english]{babel}
\usepackage[normalem]{ulem}
\usepackage{float}
\usepackage{ragged2e}

\shorttitle{Sunspot Sizes and The Solar Cycle}
\shortauthors{S. Mandal et al.}

\begin{document}

\title{Sunspot Sizes and The Solar Cycle: Analysis Using Kodaikanal White-light Digitized Data}

\author{Sudip Mandal$^{1}$,
Dipankar Banerjee$^{1,2}$}
 
\affil{$^{1}$Indian Institute of Astrophysics, Koramangala, Bangalore 560034, India. e-mail: {\color{blue}{sudip@iiap.res.in}}\\
$^{2}$ Center of Excellence in Space Sciences India, IISER Kolkata, Mohanpur 741246, West Bengal, India  \\}
  \justify
  
\begin{abstract}
   {Sizes of the sunspots vary in a wide range during the progression of a solar cycle.
 Long-term variation study of different sunspot sizes are key to better understand the underlying process of sunspot formation and their connection to the solar dynamo. Kodaikanal white-light digitized archive provides daily sunspot observations for a period of 90 years (1921-2011). Using different size criteria on the detected individual sunspots, we have generated yearly averaged sunspot area time series for the full Sun as well as for the individual hemispheres. In this paper, we have used the sunspot area values instead of sunspot numbers used in earlier studies. Analysis of these different time series show that different properties of the sunspot cycles depend on the sunspot sizes. The `odd-even rule', double peaks during the cycle maxima and the long-term periodicities in the area data are found to be present for specific sunspot sizes and are absent or not so prominent in other size ranges. Apart from that, we also find a range of periodicities in the asymmetry index which have a dependency on the sunspot sizes. These statistical differences in the different size ranges may indicate that a complex dynamo action is responsible for the generation and dynamics of sunspots with different sizes.}
\end{abstract}
\section{Introduction}
  Sunspots are the cool and dark features visible in the solar photosphere. It has a strong periodic pattern, which is popularly known as the `solar cycle' or the `sunspot cycle', where the number of sunspots increase and decrease within  11 years. Apart from the prominent 11-year period, there are various short and long-term periods present in the sunspot data. The `G-O rule' or the `odd-even rule' is one of them \citep{1948..Astron..Zh..25..18G}. According to this, the odd-numbered cycles are stronger compared to the preceding even-numbered one. This also sometime lead to a periodicity of 22 years in the sunspot number data \citep{lrsp-2010-3} which is the period of solar magnetic cycle. There are other long-term periods such as `Gleissberg cycle' with a  period of $\sim$100 years.

 Similar to the sunspot number, sunspot area also shows a 11-year period. In fact, in some cases sunspot area is considered to be a better proxy than the sunspot number \citep{2002AAS...200.5704J}. Sunspots come in different shapes and sizes and the sunspot size depend on the solar cycle phase i.e bigger spots mostly appear near the maximum of a given cycle (see \citet{2016arXiv160804665M}). Independent sunspot observations, from different observatories, show that the sunspot areas follow a log-normal distribution \citep{1988ApJ...327..451B,2005A&A...443.1061B,2016arXiv160804665M}. Cyclic variations of different sunspot sizes have been a study of great interest. Using Greenwich sunspot group number data, \citet{2012SoPh..281..827J,2016Ap&SS.361..208J} have shown the validation of the G-O rule and the Waldmeier effect for different sunspot group sizes. In an another work, using the same data, \citet{2014ARep...58..936O} have shown different correlations between the cycle amplitudes for different set of sunspot sizes.

Using the Kodaikanal white-light digitized data,  we construct different time series with different size scaling on the detected sunspots and investigate different properties of the solar cycle. Known properties of the sunspot cycle seem to hold for one range of sunspot sizes but not for all. In the following sections, we describe the data and method, followed by the results and a summary.

\section{Data Description And Method}
We use daily white-light sunspot observations, from 1921 to 2011, as recorded from the Kodaikanal observatory. In a recent initiative, the 90 years of the data has been digitized, calibrated and the extraction of the sunspots, using a semi-automated method, has also been completed recently (\citet{2016arXiv160804665M}, henceforth Paper-$\mathrm{I}$). Longitude, latitude and the area (in the units of millionth of hemisphere, $\mu$Hem) has been recorded for every detected sunspots. Using these information, different aspects of the solar cycle has been reproduced and presented in Paper-$\mathrm{I}$.\\

In this work we isolate the sunspots into five different categories according to their sizes ($S_a$ hereafter), defined as: 10$\mu$Hem$\le$S$_{a}$$<$50$\mu$Hem, 50$\mu$Hem$\le$S$_{a}$$<$100$\mu$Hem, 100$\mu$Hem$\le$S$_{a}$$<$200$\mu$Hem, 200$\mu$Hem$\le$S$_{a}$$<$500$\mu$Hem and S$_{a}$$\ge$500$\mu$Hem. These distributions are inspired from the size resolved `butterfly diagram' as plotted in Figure~12 in Paper-$\mathrm{I}$. Two small and two big sunspot ranges are chosen in order to correctly identify the switch over size range for which we notice significant changes in the cycle properties. We did not choose thresholds higher than 500$\mu$Hem as the results will be statistically insignificant due to the lesser occurrences of `very large' sunspots ($>$1000$\mu$Hem) during a particular solar cycle. We use these thresholds in the daily sunspot data and compute the yearly averaged area values from that. Here we emphasize the fact that due to different classification schemes in sunspot group identifications, the generated sunspot numbers seem to vary in different aspects \citep{2014SSRv..186...35C,2016arXiv160805261D}. 
Also the use of sunspot area instead of the sunspot number has a distinct advantage in this case. Counting sunspot number gives equal weightage to every spots whereas the area values are slightly weighted towards the big spots of the defined size range. Thus in this paper, unlike the earlier studies, we have used the area criteria, on the detected individual sunspots and calculated the total sunspot area for a particular day for the defined sunspot size range. This help us explain the results more physically.\\

\section{Results}
\begin{figure}[!htbp]
\centering
\includegraphics[width=0.49\textwidth]{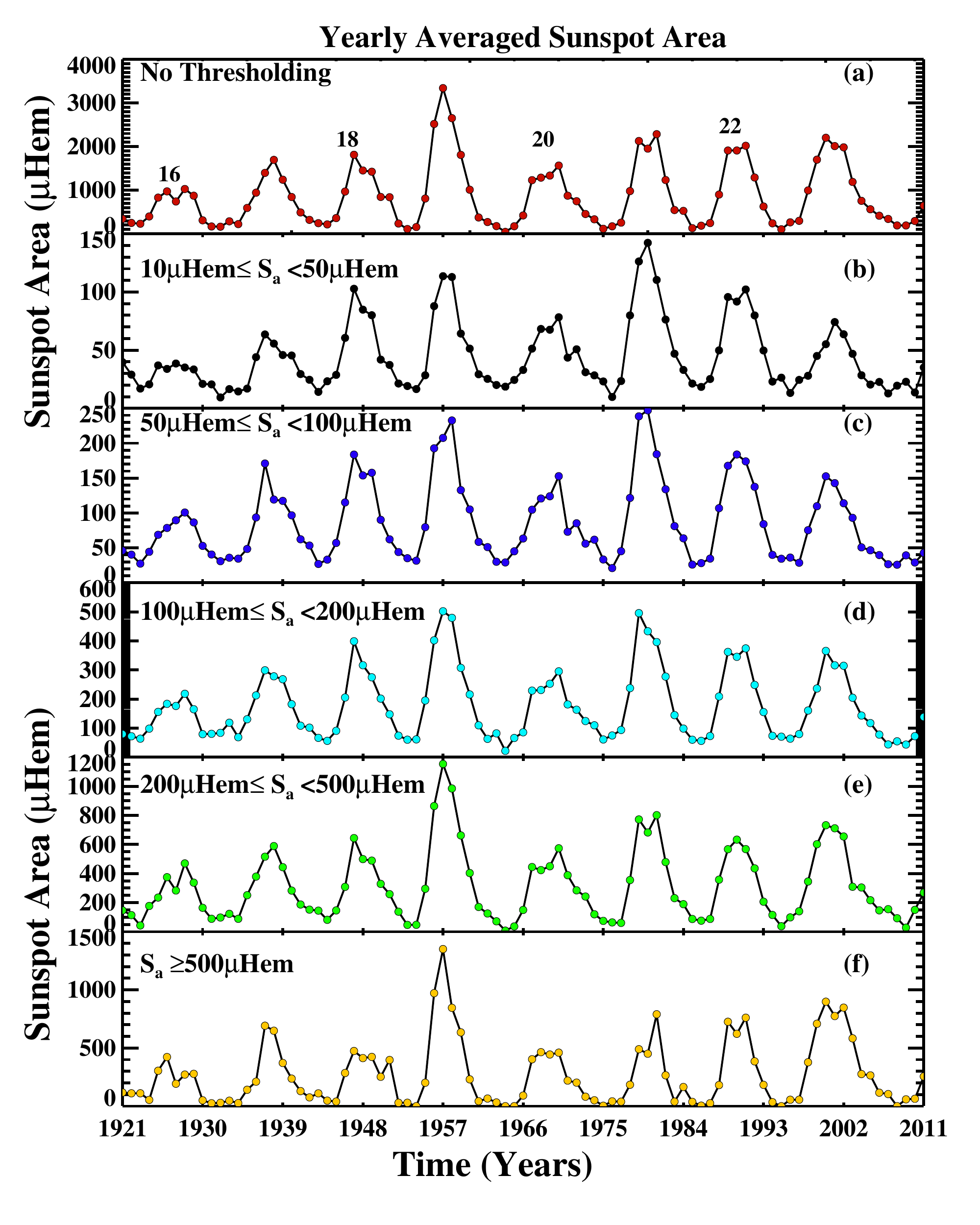}
\caption{ Yearly averaged sunspot area time series obtained for different sunspot sizes. Individual size ranges are printed on every panel.}
\label{yearly_modulation} 
\end{figure}
\subsection{Odd-Even Rule and The Double Peaks}
Different panels in Figure~\ref{yearly_modulation} show the yearly averaged sunspot area values for the period of 1921 to 2011, for different sunspot sizes. In panel \ref{yearly_modulation}.a we show area variations, for different cycles, as computed without any size restrictions. From the plot (\ref{yearly_modulation}a) we notice that the odd-numbered cycles (cycle 17, 19, 21) have higher peak values compared to the preceding even-numbered cycles (cycle 16,18,20). This is in accordance with the odd-even cycle rule \citep{1948..Astron..Zh..25..18G}. For cycle 23, the increment is minimal compared to the cycle 22. Now in panel \ref{yearly_modulation}b we plot the yearly averaged variation for the smallest sunspot size range (10$\mu$Hem$\le$S$_{a}$$<$50$\mu$Hem). Contrary to the expectation, the small sunspots do not show any significant dependence on the overall cycle strengths. In this case, the highest peak corresponds to  the 21$^{st}$ cycle whereas the strongest cycle of the last century was the 19$^{th}$ cycle. Apart form that, cycle 18 and cycle 19 also show comparable strengths for this sunspot range whereas the weaker cycles like cycle 16 and cycle 20 have relatively smaller peak heights. Also we notice a clear violation of the odd-even rule, in case of smallest sunspots, for the 23$^{rd}$ cycle as its strength is significantly lower than the previous even-numbered 22$^{nd}$ cycle. This pattern persists for the next sunspot size range of 50$\mu$Hem$\le$S$_{a}$$<$100$\mu$Hem as shown in Figure~\ref{yearly_modulation}c.\\

 However the scenario quickly changes as we move towards the mid-sized sunspot ranges (panels~\ref{yearly_modulation}c,d). The cycle strength of the 21$^{st}$ cycle goes down and become comparable to 19$^{th}$ cycle as we move towards the biggest sized sunspots. Also, the amplitude difference between the cycle 22 and cycle 23 becomes less. If we move further towards the biggest sized sunspots (panel~\ref{yearly_modulation}f), we see that the pattern matches very well with the `without thresholded' cycle variation as shown in panel~\ref{yearly_modulation}a. We  recover the `odd-even rule' again along with the cycle strengths. Thus we see that the bigger size sunspot characteristics dominates in determining the overall cycle strengths.\\
\begin{figure*}[!htbp]
\centering
\includegraphics[width=0.75\textwidth]{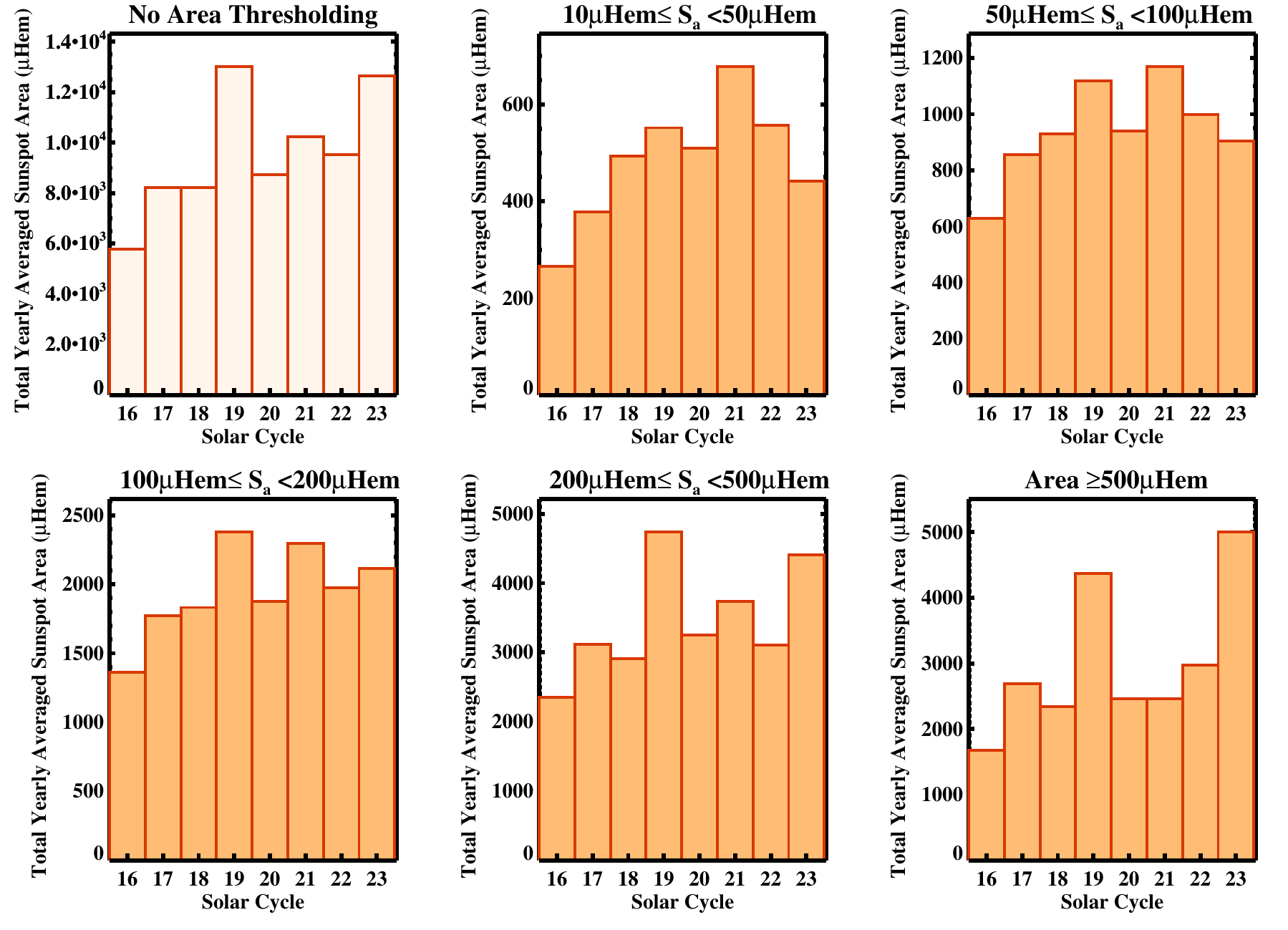}
\caption{ The variation of integrated cycle amplitudes with solar cycles. Different panels correspond to different sunspot sizes.}
\label{count_num} 
\end{figure*}

Next we focus on the occurrences of double-peaks in the solar cycles during the maximum periods \citep{2011ISRAA2011E...2G} and its relation with the sunspot sizes. In panel~\ref{yearly_modulation}a we notice that for the cycles 16, 21 and 22 there are clear signatures of double peaks. For cycles 20 and 23 there are also signature of double peaks but not as prominent as the other ones. Panels (b-f) in Figure~\ref{yearly_modulation} show that the occurrence of double peaks, for a particular cycle, is not a persistent signature in all the sunspot sizes. For an example, we see a prominent double peak in cycle 21 (panel~\ref{yearly_modulation}a), which is present for the bigger sunspot sizes (panels~\ref{yearly_modulation}e,f) but absent for smaller sunspots. In a similar example, there is a weak double peak signature for cycle 23 (panel~\ref{yearly_modulation}a) but for the biggest size range (S$_{a}\ge$500$\mu$Hem, panel~\ref{yearly_modulation}f) there is prominent double peak. In fact, in most of the cycles, for the biggest sunspot range ($S_a$$\ge$500$\mu$Hem) we see a double peak near the time of sunspot maxima. Thus we conclude that the double peaks in the solar cycle maxima occur for different sizes of the sunspots and there is no size bias i.e it may occur for small or large sunspots.

\subsection{Long-Term Variations}
 Apart from the 11-year and 22-year cycle periods, there are other long-term variations present in sunspot area data \citep{1990A&A...238..377C,1998Natur.394..552O,2002A&A...394..701K,2015LRSP...12....4H}. In order to probe this further, we use the yearly averaged sunspot area time series for different sunspot sizes. We summed the yearly averaged values for a particular cycle to produce a single number. It is thus a representation of the time averaged strength of a particular cycle. In different panels in Figure~\ref{count_num} we show the same for different sunspot sizes. For the `no thresholding' case (panel~\ref{count_num}a) we do not see any new pattern apart from regular cycle strengths i.e the maximum bar height (referring to the amplitude) corresponds to 19$^{th}$ cycle and so on.\\

  Now as we filter out the bigger sunspots i.e only keeping the smaller sunspots (panels~\ref{count_num}b,c), we see a clear pattern having a period of approximately of 10-12 cycles i.e 100-120 years. This period, in literature, is known as Gleissberg cycle \citep{2015LRSP...12....4H} which represent the amplitude modulation of the 11-year cycle period. Furthermore, the obtained patterns on panel~\ref{count_num}b and \ref{count_num}c indicate that the cycle strength for the smaller sunspots will be progressively lower for cycle 24 and cycle 25.\\

As we go for bigger sunspot sizes (panels~\ref{count_num}d,e) we notice that the pattern with the $\sim$100-120 year period disappears and the earlier trend flattens as we progressively move to bigger sunspot sizes. Here we also want to highlight the fact that for the sunspots sizes 200$\mu$Hem$\le$S$_{a}$$<$500$\mu$Hem, we notice that every odd cycle (cycle 17, 19, 21, 23) has a higher value compared to its immediate previous even cycle (cycle 16, 18, 20, 22). This is again the manifestation of the odd-even cycle rule but for the integrated cycle amplitudes. For the sunspot size $S_a$$\ge$500$\mu$Hem, the plot reveal something interesting. The height of the bar for the 23$^{rd}$ cycle is greater than that of 19$^{th}$ cycle i.e the strength of the sunspots with highest sizes, is more for the 23$^{th}$ cycle whereas the overall cycle strength is less than that of 19$^{th}$ cycle. This has happened due to extended occurrences of big sunspots during the prolonged decay phase of cycle 23 compared to cycle 19.

\subsection{North-South Asymmetry}
\begin{figure}[!htbp]
\centering
\includegraphics[trim = 5.5mm 0mm 8mm 0mm, clip,width=0.48\textwidth]{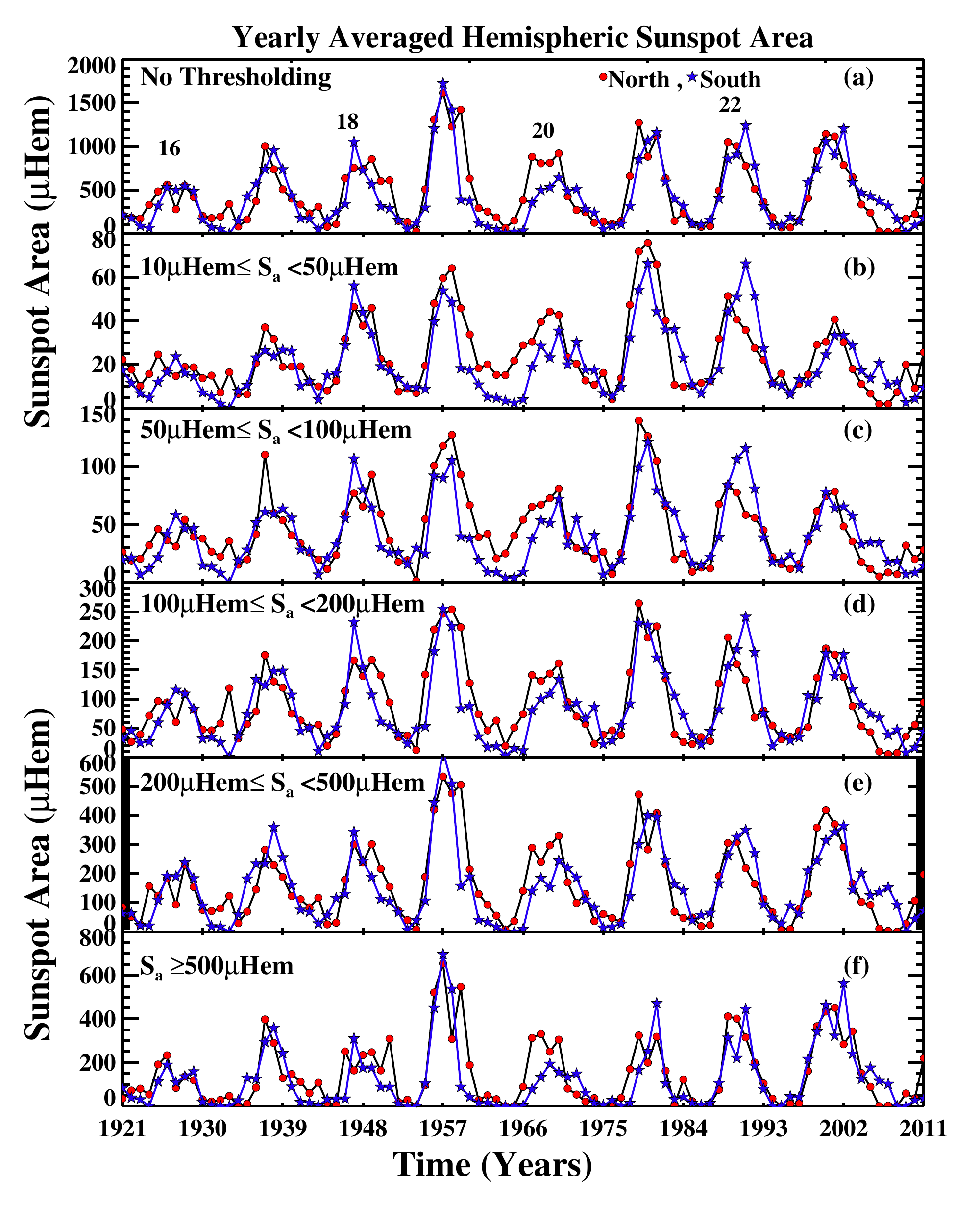}
\caption{Yearly averaged hemispheric sunspot area for different sunspot sizes. }
\label{n_s} 
\end{figure}

\begin{figure}[!htbp]
\centering
\includegraphics[trim = 5.5mm 0mm 8mm 0mm, clip,width=0.48\textwidth]{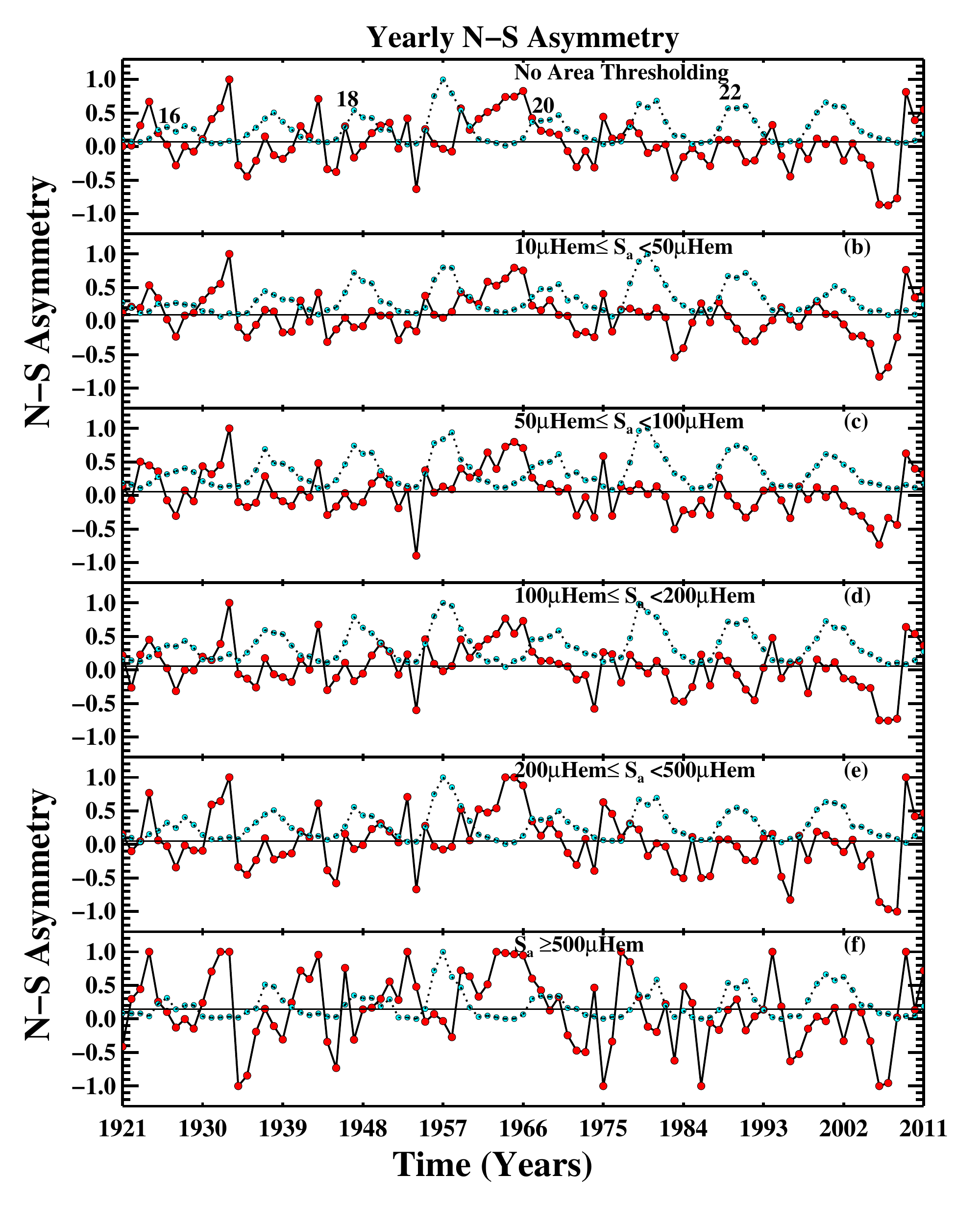}
\caption{Time evolution of the Asymmetry index, A$_{ns}$ for corresponding sunspot sizes.}
\label{coeff} 
\end{figure}


 We know that the solar activities are not symmetric in both the hemispheres and this hemispheric imbalance of the activities is known as `north-south asymmetry' or N-S asymmetry \citep{2005A&A...431L...5B,2006A&A...447..735T,2015NewA...39...55R}. Figure~\ref{n_s} shows the yearly averaged sunspot area variations, for the two hemispheres, for the defined sunspot sizes. In panel~\ref{n_s}a we show the same for the `no thresholding' case. We notice a phase difference (lead or lag) between the two hemispheres, more prominently during the epoch of cycle maxima. Apart from that we also see that the double peaks, for a given cycle, may occur for a particular hemisphere without having any counterpart of the same in the other hemisphere. As found earlier (panel~\ref{yearly_modulation}f), in this case also, most of the double peaks occur for the highest sunspot size range (panel~\ref{n_s}f).\\

 Panels~\ref{n_s}(b-f) show the hemispheric yearly sunspot for different sunspot sizes. A closer inspection reveal that for different sunspot sizes, one hemisphere dominates compared to the other and the vice-verse occur for different size ranges. As an example, for cycle 19, the northern hemisphere dominates over the southern for the small sunspots whereas for the progressively larger sunspots the opposite behavior is seen i.e south dominates over north. In contrast, for cycle 22, the south dominates over the north for small sunspots but gradually the difference minimizes as we reach towards bigger sunspots. Though we must emphasize that there are cycles for which we do not see any big change in the two hemisphere for any sunspot size range. Cycle 16, 17, 23 are the examples of such cases.\\

  In order to quantify the asymmetry between the two hemispheres, we use the `north-south asymmetry index' (A$_{ns}$) as (S$_{an}$-S$_{as}$)/(S$_{an}$+S$_{as}$), where S$_{an}$ and S$_{as}$ are the yearly averaged sunspot area values for the northern and the southern hemispheres respectively. We plot the time variation of A$_{ns}$ for different sunspot sizes in different panels of Figure~\ref{coeff}. In a first glance we notice that the variation of A$_{ns}$ is very smooth for the smallest sunspot sizes (panel~\ref{coeff}b), as compared to any other size range (panels~\ref{coeff}c-f). This can be understood from the definition of A$_{ns}$ which varies between $\pm$1, depending upon the presence of at least one single sunspot in any of the hemispheres and absence of any spot in the other hemisphere.  Since, the smaller sunspots occur more frequently at the time of solar minima compared to the appearance of a big sunspot, so the variation of A$_{ns}$ is relatively smoother in the former case.  We also notice (from panel~\ref{coeff}a) that at the maximum phase of cycle 21, A$_{ns}$ has values very close to zero as in the case of cycle 23. However, at the cycle 22 maximum A$_{ns}$ attains a maximum value of $\approx$-0.5. This is consistent with the results, as obtained by \citet{2005MNRAS.362.1311J}. This scenario changes a little as we move from small to large sunspot sizes. There is a hint of long-term variation in the asymmetry index. For the largest sized sunspots (panel~\ref{coeff}f), the northern hemisphere dominates in the cycle minima till cycle 20, after which southern hemisphere dominates. This trend is visible in panel~\ref{coeff}e also. 

\subsubsection{Periodicities in the N-S Asymmetry}

 Since the solar cycle has a dominant period of 11 years, it is expected that the N-S asymmetry index will also have some periodicity. There have been a lot of work done in order to probe this periodicity \citep{1993A&A...274..497C,2005A&A...431L...5B,2009NewA...14..133C,2015NewA...34...54J,2015NewA...39...55R}. Here we revisit the periodicity in the N-S asymmetry for different sunspot sizes.\\

\begin{figure*}[!htbp]
\centering
\includegraphics[trim = 12mm 0mm 105mm 0mm, clip,width=0.22\textwidth,angle=90]{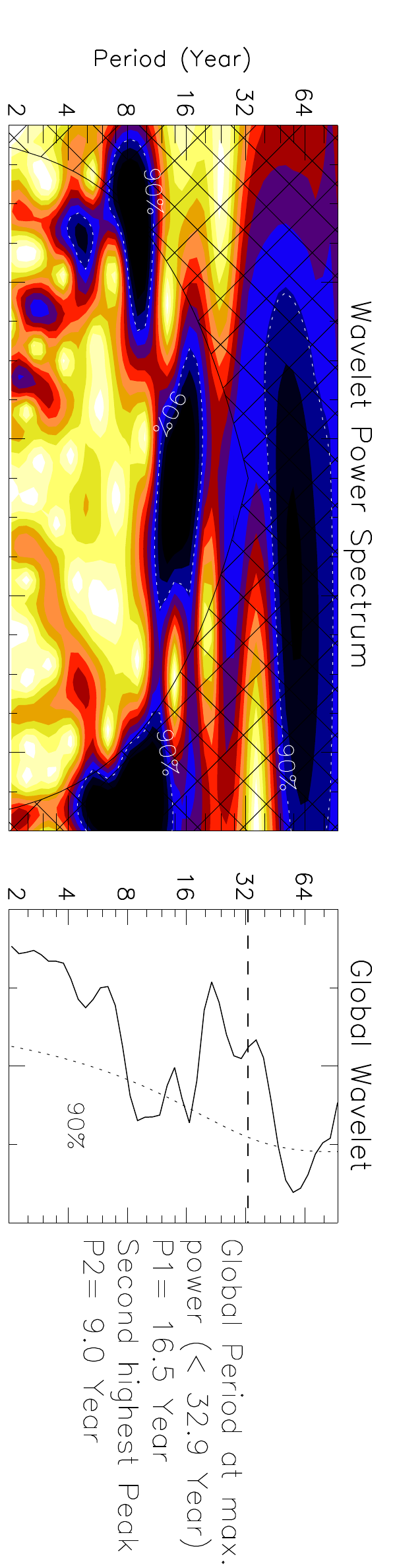}
\includegraphics[trim = 12mm 0mm 113mm 0mm, clip,width=0.19\textwidth,angle=90]{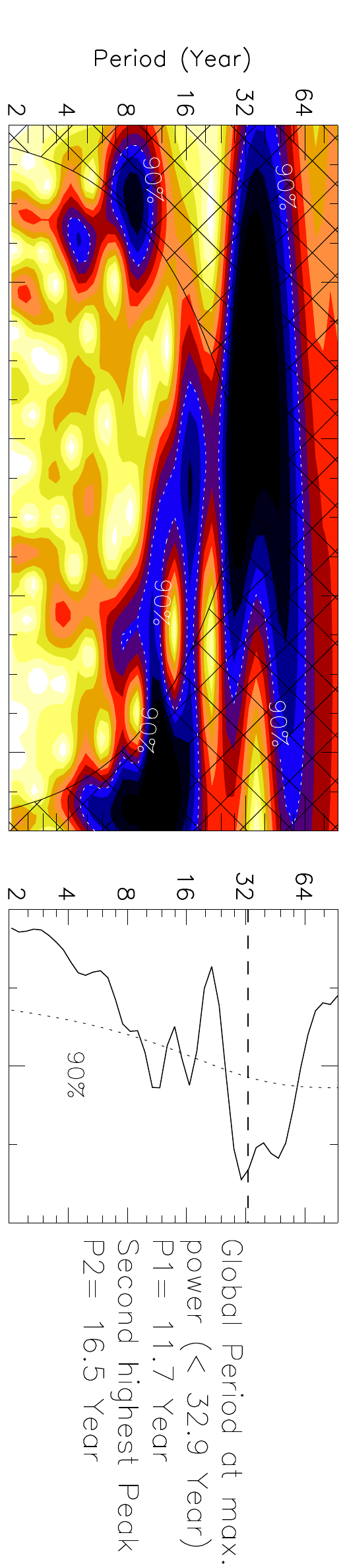}
\includegraphics[trim = 12mm 0mm 113mm 0mm, clip,width=0.19\textwidth,angle=90]{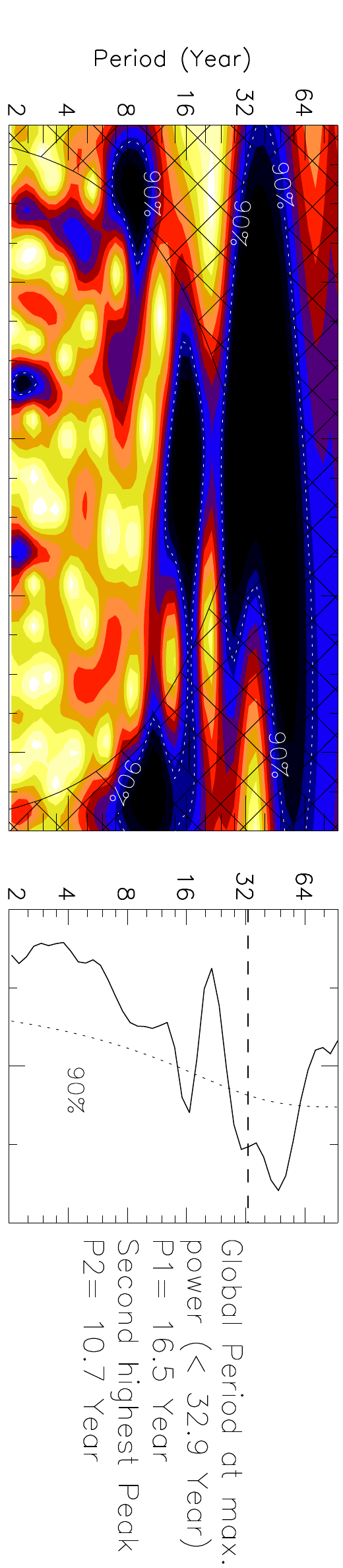}
\includegraphics[trim = 12mm 0mm 113mm 0mm, clip,width=0.19\textwidth,angle=90]{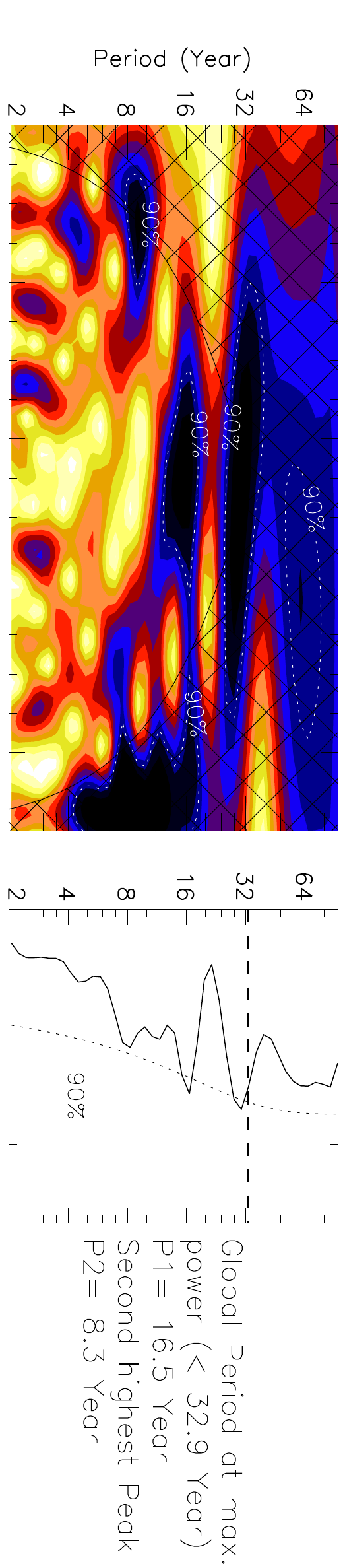}
\includegraphics[trim = 12mm 0mm 113mm 0mm, clip,width=0.19\textwidth,angle=90]{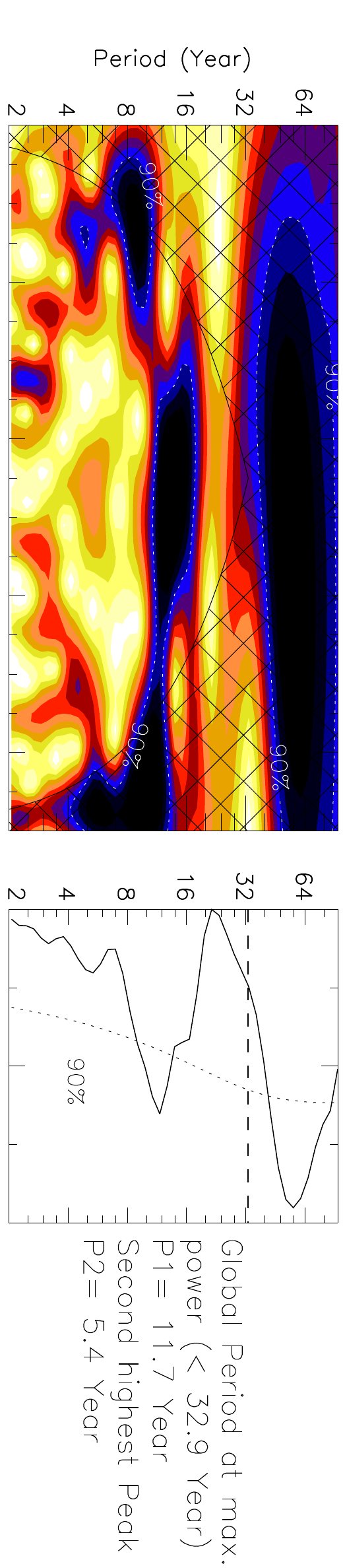}
\includegraphics[trim = 0mm 0mm 113mm 0mm, clip,width=0.232\textwidth,angle=90]{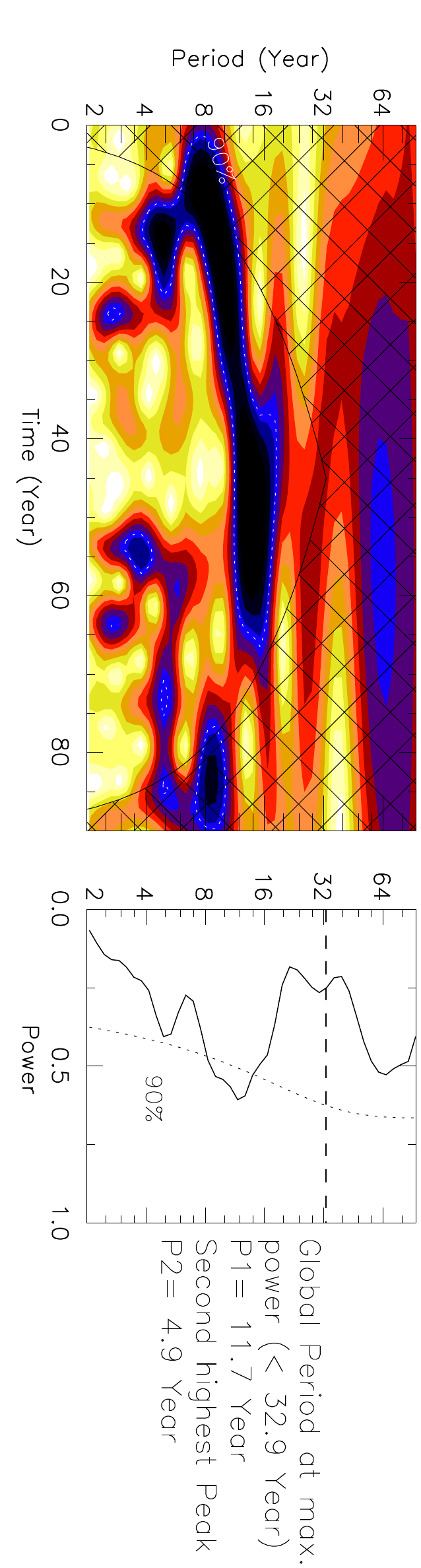}

\caption{[Top to bottom]: Results of wavelet analysis corresponding to  the light curves of A$_{ns}$, as shown in panels~\ref{coeff}(a-f) respectively. The periods with the maximum significant powers are listed after the wavelet power spectrum (left panel) and the global wavelet plot (middle panel).} 

\label{ns_period}
\end{figure*}

  We use the A$_{ns}$ curves, shown in different panels in Figure~\ref{coeff}, and use the wavelet analysis in order to obtain the periodicities in A$_{ns}$. Results of the wavelet analysis on the different A$_{ns}$ curves is shown in Figure~\ref{ns_period}. The global wavelet power (time averaged power as shown in `wavelet power spectrum panel') is calculated along with a significance level 90\%. This significance level is obtained for the white noise \citep{1998BAMS...79...61T} and contours are overplotted on top of the wavelet spectrum in order to highlight the region above the confidence level. In the global wavelet panel, a horizontal dashed line indicate the maximum measurable period due to the `cone of influence' (COI) (which arises due to the edge effect and represented by the cross-hatched region marked in the wavelet power spectrum panel).\\

 The top panel in Figure~\ref{ns_period} show the period obtained from the `no thresholding' case where we see that the dominant period is 16.5 years \citep{2015NewA...34...54J} and the second highest period is 9 years \citep{2009NewA...14..133C}. Wavelet spectrum and the corresponding periods obtained for different sunspot sizes are shown in consecutive panels in Figure~\ref{ns_period}. We notice that the dominant periods for the small sunspot sizes (sizes from 10$\mu$Hem$\le$S$_{a}$$<$100$\mu$Hem), are 10.7 years, 11.7 years \citet{1993A&A...274..497C} and 16.5 years. Now as we move towards bigger sized sunspots (100$\mu$Hem$\le$S$_{a}$$<$500$\mu$Hem), shorter periodicities are observed to appear.  Apart form the 10.7 years and 16.5 years, we now have periodicities of 8.3 years and 5.4 years. For the biggest sized sunspots ($\ge$500$\mu$Hem), we even get as small period as 4.9 years apart from the dominant 11.7 years. We also see that there are two other periods at $\sim$30 years and $\sim$46 years present for different sunspot sizes. However these two periods are beyond the faithful detection level determined by the COI (this arises due to the duration of the time series). Though it is worth mentioning here that a period of $\approx$44 years , double of the solar magnetic cycle, has been reported earlier from the N-S asymmetry \citep{2005A&A...431L...5B}.\\

\section{Summary and Conclusions}

 Different sunspot sizes manifests different distinct properties on shorter and longer time scales compared to the dominant 11-years period. In our analysis we have used sunspot area values, which turns out to be a better proxy than sunspot number for this type of studies. Below we summarize the main results found from the analysis:\\

$\bullet$  Our analysis show that the overall pattern of a cycle is primarily dictated by the bigger size sunspots. Small sized sunspots  show no clear correlation with the overall cycle strengths. We found cycle 21 to be the strongest cycle considering the small sunspots only. At the same time we notice that, for this cycle, the cycle strength considering the big spots decreases gradually. This is also true for cycle 18 and cycle 22. For 19$^{th}$ the case gets reversed.
Thus this supports the idea of a dynamo mechanism where small sunspots are the fragmented part of the bigger sunspots.\\

$\bullet$ For cycle 23, the odd-even rule gets violated, specially for the small sized sunspots. However, for the bigger spots, the anomaly seem to reduce. Also the double peaks at the solar maxima are not found in all the sunspot sizes. Though not prominent, but the trend show that the occurrences of double peaks is maximum for the biggest sunspot sizes, whereas occasionally it shows up for the small sunspots. \\

$\bullet$ A clear pattern of $\sim$120-130 years period has been found for the small sunspot sizes. The observed trend also implies that the cycle strengths of cycle 24 and cycle 25, for the small sunspot, will be weaker than that of cycle 23. Here we notice that the pattern disappears for the big sunspots indicating that on a long-term basis (longer than the 11-year period) also the two size scales have two different trends.\\ 

$\bullet$ Hemispheric asymmetry is found to be different for different sunspot sizes. The double-peak behavior, for a particular cycle, also shows a hemispheric dependence.  Analyzing the asymmetry index (A$_{ns}$) times series, we found a dominant period of 9 years, $\approx$12 years and 16.5 years in most of the sunspot sizes. Apart form that, we obtained smaller periods like 8.3 years, 5.4 years and 4.9 years for the bigger sized sunspots. The presence of periods $\approx$5 years can be a manifestation of the asymmetric nature of the solar cycle.\\

There is currently no understanding of the physical reason for a possible sunspot size dependence on the solar dynamo. Sunspot area distributions, for the big and the small sunspot sizes, have been found to be distinctively different \citep{1988ApJ...327..451B}. Results from our analysis also indicate such anomalies for big and small sunspots. We thus conjecture that these phenomena can be explained by a dynamo formulation with two components, one directly connected to the global component of the dynamo (and the generation of bipolar active regions), and the other with the small-scale component of the dynamo (and the fragmentation of magnetic structures due to their interaction with turbulent convection)  \citep{2015ApJ...800...48M}.\\

To conclude, we have analyzed daily sunspot data, obtained from Kodaikanal, in order to investigate the sunspot size dependence of various solar cycle features. We found distinct signatures present for the small, medium and big sized sunspot area time series which may indicate a complex dynamo operating differently on different size scales. Independent studies using other solar proxies and other data set will help us to understand the underlying mechanism responsible for these phenomena. 

\section{Acknowledgment}
{The authors would like to thank the referee for his/her valuable suggestions which helped us for a better presentation of the paper. We would also like to thank the Kodaikanal facility of Indian Institute of Astrophysics, Bangalore, India for proving the data. This data is now available for public use at \url{http://kso.iiap. res.in/data}.}\\

 \bibliographystyle{apj}

\begin{thebibliography}{25}
\expandafter\ifx\csname natexlab\endcsname\relax\def\natexlab#1{#1}\fi

\bibitem[{{Ballester} {et~al.}(2005){Ballester}, {Oliver}, \&
  {Carbonell}}]{2005A&A...431L...5B}
{Ballester}, J.~L., {Oliver}, R., \& {Carbonell}, M. 2005, \aap, 431, L5

\bibitem[{{Baumann} \& {Solanki}(2005)}]{2005A&A...443.1061B}
{Baumann}, I., \& {Solanki}, S.~K. 2005, \aap, 443, 1061

\bibitem[{{Bogdan} {et~al.}(1988){Bogdan}, {Gilman}, {Lerche}, \&
  {Howard}}]{1988ApJ...327..451B}
{Bogdan}, T.~J., {Gilman}, P.~A., {Lerche}, I., \& {Howard}, R. 1988, \apj,
  327, 451

\bibitem[{{Carbonell} \& {Ballester}(1990)}]{1990A&A...238..377C}
{Carbonell}, M., \& {Ballester}, J.~L. 1990, \aap, 238, 377

\bibitem[{{Carbonell} {et~al.}(1993){Carbonell}, {Oliver}, \&
  {Ballester}}]{1993A&A...274..497C}
{Carbonell}, M., {Oliver}, R., \& {Ballester}, J.~L. 1993, \aap, 274, 497

\bibitem[{{Chang}(2009)}]{2009NewA...14..133C}
{Chang}, H.-Y. 2009, \na, 14, 133

\bibitem[{Charbonneau(2010)}]{lrsp-2010-3}
Charbonneau, P. 2010, Living Reviews in Solar Physics, 7

\bibitem[{{Clette} {et~al.}(2014){Clette}, {Svalgaard}, {Vaquero}, \&
  {Cliver}}]{2014SSRv..186...35C}
{Clette}, F., {Svalgaard}, L., {Vaquero}, J.~M., \& {Cliver}, E.~W. 2014, \ssr,
  186, 35

\bibitem[{{Dudok de Wit} {et~al.}(2016){Dudok de Wit}, {Lef{\`e}vre}, \&
  {Clette}}]{2016arXiv160805261D}
{Dudok de Wit}, T., {Lef{\`e}vre}, L., \& {Clette}, F. 2016, ArXiv e-prints

\bibitem[{{Georgieva}(2011)}]{2011ISRAA2011E...2G}
{Georgieva}, K. 2011, ISRN Astronomy and Astrophysics, 2011, 437838

\bibitem[{{Gnevyshev} \& {Ohl}(1948)}]{1948..Astron..Zh..25..18G}
{Gnevyshev}, M.~N., \& {Ohl}, A.~I. 1948, Astron. Zh., 25, 18

\bibitem[{{Hathaway}(2015)}]{2015LRSP...12....4H}
{Hathaway}, D.~H. 2015, Living Reviews in Solar Physics, 12

\bibitem[{{Javaraiah}(2005)}]{2005MNRAS.362.1311J}
{Javaraiah}, J. 2005, \mnras, 362, 1311

\bibitem[{{Javaraiah}(2012)}]{2012SoPh..281..827J}
---. 2012, \solphys, 281, 827

\bibitem[{{Javaraiah}(2015)}]{2015NewA...34...54J}
---. 2015, \na, 34, 54

\bibitem[{{Javaraiah}(2016)}]{2016Ap&SS.361..208J}
---. 2016, \apss, 361, 208

\bibitem[{{Jordan} \& {Garcia}(2002)}]{2002AAS...200.5704J}
{Jordan}, S.~D., \& {Garcia}, A.~G. 2002, in Bulletin of the American
  Astronomical Society, Vol.~34, American Astronomical Society Meeting
  Abstracts \#200, 737

\bibitem[{{Krivova} \& {Solanki}(2002)}]{2002A&A...394..701K}
{Krivova}, N.~A., \& {Solanki}, S.~K. 2002, \aap, 394, 701

\bibitem[{{Mandal} {et~al.}(2016){Mandal}, {Hegde}, {Samanta}, {Hazra},
  {Banerjee}, \& {B}}]{2016arXiv160804665M}
{Mandal}, S., {Hegde}, M., {Samanta}, T., {Hazra}, G., {Banerjee}, D., \& {B},
  R. 2016, ArXiv e-prints

\bibitem[{{Mu{\~n}oz-Jaramillo} {et~al.}(2015){Mu{\~n}oz-Jaramillo},
  {Senkpeil}, {Windmueller}, {Amouzou}, {Longcope}, {Tlatov}, {Nagovitsyn},
  {Pevtsov}, {Chapman}, {Cookson}, {Yeates}, {Watson}, {Balmaceda}, {DeLuca},
  \& {Martens}}]{2015ApJ...800...48M}
{Mu{\~n}oz-Jaramillo}, A., {et~al.} 2015, \apj, 800, 48

\bibitem[{{Obridko} \& {Badalyan}(2014)}]{2014ARep...58..936O}
{Obridko}, V.~N., \& {Badalyan}, O.~G. 2014, Astronomy Reports, 58, 936

\bibitem[{{Oliver} {et~al.}(1998){Oliver}, {Ballester}, \&
  {Baudin}}]{1998Natur.394..552O}
{Oliver}, R., {Ballester}, J.~L., \& {Baudin}, F. 1998, \nat, 394, 552

\bibitem[{{Ravindra} \& {Javaraiah}(2015)}]{2015NewA...39...55R}
{Ravindra}, B., \& {Javaraiah}, J. 2015, \na, 39, 55

\bibitem[{{Temmer} {et~al.}(2006){Temmer}, {Ryb{\'a}k}, {Bend{\'{\i}}k},
  {Veronig}, {Vogler}, {Otruba}, {P{\"o}tzi}, \&
  {Hanslmeier}}]{2006A&A...447..735T}
{Temmer}, M., {Ryb{\'a}k}, J., {Bend{\'{\i}}k}, P., {Veronig}, A., {Vogler},
  F., {Otruba}, W., {P{\"o}tzi}, W., \& {Hanslmeier}, A. 2006, \aap, 447, 735

\bibitem[{{Torrence} \& {Compo}(1998)}]{1998BAMS...79...61T}
{Torrence}, C., \& {Compo}, G.~P. 1998, Bulletin of the American Meteorological
  Society, 79, 61

\end{thebibliography}

\end{document}